\documentclass[letter]{aa}  
\usepackage{graphicx}
\usepackage{epsfig,natbib,lscape,supertabular,txfonts}
\begin{document}
   \title{New light on the formation and evolution of bars}
   \subtitle{Trends in the stellar line-strength indices distribution inside the bar region \thanks{Based on observations obtained at Siding Spring Observatory (RSAA, ANU, Australia)}}

     \author{I. P\'erez \inst{1}\thanks{Veni
          Fellow}\fnmsep\inst{2}\thanks{Associate Researcher}\and
          P. S\'anchez-Bl\'azquez\inst{3}\thanks{Marie Curie}\and A. Zurita\inst{2}\thanks{Retorno J.A Fellow}}


   \institute{Kapteyn Astronomical Institute, University of Groningen, the Netherlands
\\ email:isa@astro.rug.nl \and Departamento de F\'isica te\'orica  y del Cosmos, Universidad de Granada, Spain\\ email:azurita@ugr.es \and Centre For Astrophysics, University of Central Lancashire, UK
\\email:psanchez-blazquez@uclan.ac.uk }

   \date{Received September 15, 1996; accepted March 16, 1997}


  \abstract
   {}
   {The aim is to study the stellar content of the bar region to constrain its formation and evolution.}
   {Line-strength indices in the bar region of a sample of 6 barred galaxies were employed to derive age and metallicity gradients along the bars using stellar population models.}
   {We find clear radial gradients in the line-strength indices for all the galaxies. We find positive gradients within the bar region in the metal indices in four of the six galaxies, and opposite trends in the other two. These latter two galaxies are classified as SAB and they present exponential bar light profiles. For all the galaxies we find a positive gradient in the Balmer indices. There is a clear correlation between the position of morphological features inside the bar region with changes in the slope and value of the indices, which indicate, using stellar population analysis, changes in the stellar populations. Therefore, it seems that the bar regions show a gradient in both age and metallicity, changing radially to younger and more metal rich populations for all the galaxies except for those two with exponential profiles.}
   {}

   \keywords{Galaxies: abundances -- Galaxies:formation  -- Galaxies: evolution }

   \maketitle
%

\section{Introduction}

The importance of bars in  galaxies as mechanisms to transfer angular
momentum and matter on large scales within the galaxy is
indisputable~\citep[][and references therein]{kormendy}. Bars are
thought to be formed through spontaneous disk instabilities or by
instabilities produced during galaxy encounters; the details of bar
formation and their subsequent evolution remains a matter of
debate. Until now, only detailed analyses of the gas-phase abundances
along bars have been undertaken, from observations of emission lines of
H{\small II} regions. Studies of the gas-phase abundances provide us with a 
present-day
snapshot of the interstellar medium abundance. These studies have
shown that there is little variation in the 
chemical abundances in the gas-phase
along the bar \citep{martin97,martin99}. The study of the stellar ages
and metallicities, on the other hand, gives us 'archaeological' clues
as to the formation and evolution of the bar. Furthermore, the evolution
of the two components, gas and stars, suffers from very different
evolutionary processes; the gas is mainly dominated by the
gravitational torque of the non-axisymmetric mass component, while the
evolution of the stellar component is mainly affected by different 
orbital mixing
\citep{binney}, so one would not expect the same abundance trends in
both components. \cite{friedli98} predicted, using N-body simulations of
bars with pre-existing exponential abundances, a null evolution of the
stellar abundance profile (although with a decrease in the mean
metallicity) while the gas abundance profile flattened rapidly. As it
has been mentioned before, the gas-phase abundance distribution predicted by
\cite{friedli98} has been shown to agree with the observational data
of abundances in H{\small II} regions.  As for the stellar component,
there has not been any study addressing the stellar population along bars
using spectroscopic information. Recently, \cite{moorthy} have
presented results for the stellar populations of bulges of spiral
galaxies, some of which are barred galaxies, finding a difference between
bulges of barred and un-barred galaxies.

To date, it has been difficult to obtain radial abundance
distributions along bars because, in order to obtain reliable stellar
abundances and ages, very high signal-to-noise (S/N) spectra are
necessary, and although bars are high surface brightness structures,
this still implies long integration times on medium-size
telescopes. We have carried out a project to obtain high S/N data
along the bar major axis of a sample of barred galaxies.  This is the
first detailed study of stellar line indices along the bar region. In
this letter, we show clear evidence for line-strength index gradients
along the bars and a correlation between the position of morphological
features inside the bar region
(bulge/inner-disk/inner-bar/primary-bar) and breaks in their
respective stellar populations.

\section{Observations and data reduction}

\subsection{Sample selection}


We have selected barred galaxies from the Third Reference Catalogue of
bright galaxies (RC3)~\citep{vaucouleurs}, with the following
criteria; to be southern ($\delta < 0^{\circ}$), classified as barred
and with inclinations between 10$^{\circ}$ and 50$^{\circ}$, and
nearby ($cz$~$\le$~2500~km~s$^{-1}$) to be able to properly resolve
the bar, with morphological types earlier than SBb to avoid
morphological dependency of the results (in this first approach) and
to ensure that the requested S/N is reached, since early-type galaxies
have higher bar surface brightness, crucial for the line-strength
determination. We finally observed six galaxies (names and
morphological types are indicated in Fig.~\ref{fig:indices}). It is by
no means a statistically complete sample, but it is a starting point
in the study of line-strength indices in bars.

\subsection{Observations and data reduction}

We obtained long--slit spectra of six barred galaxies with the Double
Beam Spectrograph (DBS) on the 2.3m telescope at Siding Spring
Observatory (SSO) during February 2006. SITe $1752\times532$ CCD
detectors were used. The gratings employed were the 600B and the 600R
for the blue and red arms, respectively, with a slit-width of
2~arcsec. This set-up gives a dispersion of 1.1~\AA/pixel for the blue
arm and 1.09~\AA/pixel for the red arm in the wavelength intervals
3892--5814 \AA\ and 5390--7314 \AA\ respectively, giving a velocity
dispersion resolution $\Delta\sigma~\approx~$130km~s$^{-1}$. The slit
was placed along the bar major axis. The slit length is 7~arcmin,
which allows one to study nearby galaxies in detail, while still
providing sufficient sky coverage. 
The position angle was derived using the DSS images of the sample galaxies. 
Comparison arc lamp
exposures were obtained for wavelength calibration. Spectrophotometric standards were observed with a slit width
of 6 arcsec. Additionally, we observed  11 G-K stars from the Lick/IDS
library to be used as templates for velocity dispersion ($\sigma$)
measurements, as well as to transform our line-strength indices to the
Lick system. The total integration time for each of the galaxies was
typically 3 hours.

The data presented in this letter refers to the blue range alone. All
the spectra were reduced using standard IRAF\footnote{IRAF is
distributed by the National Optical Astronomy Observatories, which is
operated by the Association of Universities for Research in Astronomy,
Inc. (AURA) under cooperative agreement with the National Science
Foundation.} routines. Overscan and bias were subtracted from all the
frames. No dark subtraction was done due to the low dark current of
the chips used. Flatfielding correction was achieved to the 2\%
level. The relatively small, compared to the slit length, size of the
galaxies allowed for accurate sky subtraction for each frame prior to
combining the reduced spectra. A first order polynomial was fit along
the spatial direction and then subtracted from the
frames. Spectrophotometric standards were used to relatively flux
calibrate the spectra.  For each fully reduced galaxy frame, a final
frame was created by extracting spectra along the slit, binning in the
spatial direction  to guarantee a minimum signal-to-noise ratio of 20
per \AA~in the spectral region of Mgb.

\section{Line-strength indices}
\label{indices}
In order to analyse the stellar population of the bars, spectral
indices of the Lick/IDS system~\citep{1985ApJS...57..711F} were
measured and are presented in Fig.~\ref{fig:indices}.  We adopted the definition of~\cite{1998ApJS..116....1T}
and~\cite{1997ApJS..111..377W} for the high-order Balmer lines
H$\delta$ and  H$\gamma$. The Fe3n index derived here is a combination of three prominent Fe lines (
Fe4383+Fe5270+Fe5335)/3 ), similar to the Fe3 index defined
by~\cite{Kuntschner}, with Fe4383 instead of Fe5015.

To compare between different regions of the galaxy and to compare with
other galaxies, indices have to be measured at the same resolution.
As to allow flexibility in the model selection and to
potentially  compare with the measurements of other galaxy components
from other studies, we decided to follow this approach. After
broadening to the Lick resolution,  line-strength indices were
measured using the INDEX routine~\citep{1999PhDT........12C}.

To derive errors in the line-strength indices, we followed a similar
approach as  in~\cite{2006astro.ph..6642K}. This method consists in
calculating the errors in the mean flux within a passband using noise
spectra derived by subtracting a best-fit model to the galaxy
spectra. To derive the best-fit we used the OPTEMA algorithm described
by~\cite{1993PhDT.......172G}.
We used the synthetic spectral energy distributions  provided by
Vazdekis et al. (2007). Once the error in the mean flux of each
bandpass is calculated, the formulae derived in~\cite{cardiel98} were
used to calculate the errors in the indices. As the values of the
indices depends on the Doppler broadening, the velocity dispersion
along the radii was also calculated. To derive $\sigma$  and radial
velocities, we use the MOVEL and  OPTEMA algorithms described
by~\cite{1993PhDT.......172G}.  To build an optimal template we use
stars observed with exactly the same instrumental configuration as the
galaxies.  To correct the indices for the velocity dispersion
broadening we followed the standard procedure of deriving correction
curves from artificially broadened spectra. We derive, for each
galaxy, a different curve using the best-fit spectra obtained in the
derivation of the errors.  The reason to derive a polynomial for each
galaxy  is that the broadening correction depends on the strength of
the indices~\citep{1998PhDT........24K}, and by adopting a single
polynomial for all the galaxies, artificial trends between indices and
$\sigma$ can be introduced~\citep{2006astro.ph..6642K}.

To correct the indices for any nebular emission contribution we
used GANDALF~\citep{sarzi} which is a simultaneous emission and
absorption lines fitting algorithm. We have marked all the points with
detected emission (see Fig.~\ref{fig:indices}), where the detection
threshold has been calculated with the prescriptions given
in~\cite{sarzi}. The sensitivity vary from 0.3~\AA~in the internal
parts to 0.6~\AA~in the external parts of the galaxy. When the
emission completely fills all the absorption Balmer lines, the
correction is more uncertain. For those cases, we have used different symbols in Fig.~\ref{fig:indices}.

We used TMB03,~\cite{2003MNRAS.344.1000B} and Vazdekis et al. (2007)
models to derive simple stellar population ages and metallicities
using the indices H$\beta$ and [MgbFe]$'$ (TMB03).  
We also derived [$\alpha$/Fe] using
Fe4383 and Mgb by fixing the ages obtained in the [MgbF]$'$
vs. H$\beta$ diagram and the TMB03 models.

\section{Location of the morphological features}

In order to locate the radius of changes in morphological structure,
ellipse fitting to the light distribution was performed using the IRAF
task ELLIPSE. All the galaxies showed point-like nuclear regions, so
the centre was fixed using the coordinates obtained by fitting a
Gaussian to the nucleus. The position angle (PA) and the ellipticity
($e$) were left as free parameters in the fitting. Changes in both
ellipticity and PA were considered as candidate structures, then
visual inspection of the images was performed to see if there was any
anomaly. The analysis is similar to that described in
\cite{erwin}. The images used for the analysis come from different
sources, all the analysis was done with 2MASS data and also a
different band in the cases were there were published images. The
following sources were used for each of the galaxies: for NGC~3081 the
HST image of the nuclear region in the F606W band \citep{malkan}, for
NGC~1433 the $R$-band by \cite{hameed}, and for NGC~5101 $B$-band
image by \cite{eskridge}. We compared the results in the different
bands finding good agreement in the position of the morphological
features.

For three out of the six galaxies (NGC~1433, NGC~3081, and NGC~2217)
there are published data on the radii of morphological structures
within the inner region \citep{erwin}. There is good agreement
(differences of $\approx$ 5\%) between our analysis and the
values found by \cite{erwin}.

\section{Results and summary}

\begin{figure*}
\begin{center}
\vspace{0cm}
\hspace{0cm}\psfig{figure=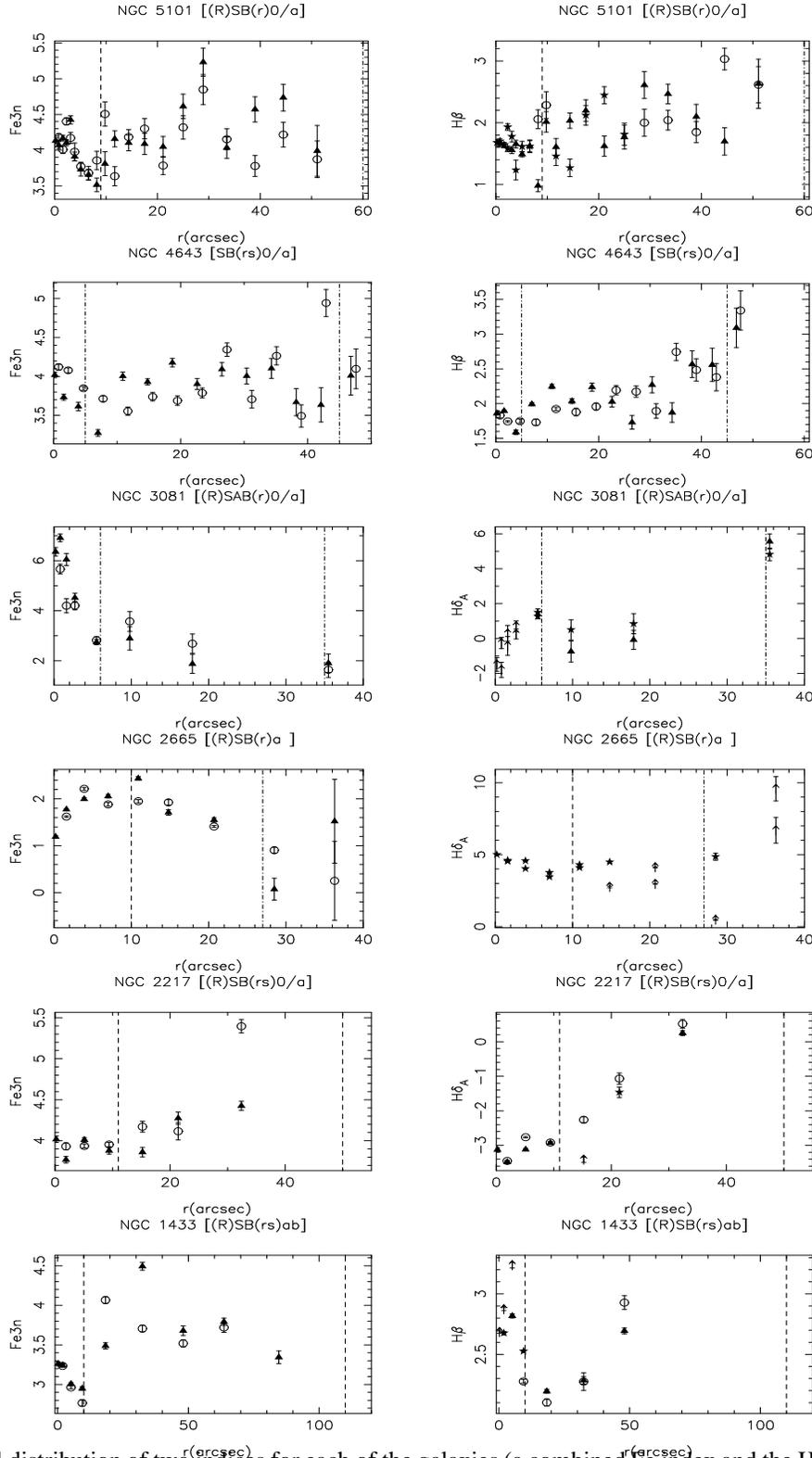,width=17.0cm}
\vspace{0mm}
\caption {Radial distribution of two indices for each of the galaxies
(a combined Fe index and the H$_{\beta}$ index or the H$_\delta$ image
depending on the emission in the bin, see text). The open circles and
the filled triangles represent both sides of the bar. We have marked
all the points with detected emission with a star, where the detection
threshold has been calculated with the prescriptions given
in~\cite{sarzi}. When the emission completely fills all the absorption
Balmer lines, the correction is more uncertain. For thoses cases, we
have  marked the points with an arrow. The vertical dashed lines
represent the end-point of morphological features, meaning the inner
bar or nuclear disk and the primary bar. Notice the general change in
the indices for the two regions and the gradients in both.}
\label{fig:indices}
\end{center}
\end{figure*}

Figure~\ref{fig:indices} shows the radial trends of two indices for
each of the galaxies, a metal and a Balmer index. The other Balmer
indices follow the same trends shown in Fig.~\ref{fig:indices}. The Mg
indices in some cases differ from the Fe indices, following opposite
trends. However, one has to bear in mind that raw line indices are
affected by the age--metallicity degeneracy. Therefore, to disentangle
age and metallicity effects from the indices, one requires the use of
stellar population models. In the cases where H$\beta$ is more
affected by emission H$\delta$ was chosen, since it is less
affected by the emission lines (see Section 3 about emission
correction). One should be careful deriving the Balmer indices affected
with emission since the derived values are sensitive to the emission
correction (increasing the value after the correction). We have tried
several methods (even no correction) and the trends do not change. In
Fig.~\ref{fig:indices} one can clearly see a different trend of the
radial distribution of the indices with a change at the radius of the
inner structure (in Fig.~\ref{fig:indices}, the most inner vertical
line).

For clarity, we divide the results of the radial
distribution in: a) the inner structure, up to the most inner
vertical line in Fig.~\ref{fig:indices}, and b) the bar region, between
the two vertical lines in Fig.~\ref{fig:indices}.

a) Inner region. There seems to be a  gradient of the metal and Balmer
lines within the inner structure, without taking into account the
central points. These very central points are possibly dominated by
the bulge, and show a different behaviour from the bar and the inner
structure. Both sides of the bar show similar trends, reinforcing the results.

b) Bar region. Within the bar region the slope changes for both the
metal and the Balmer indices. For NGC~5101 and NGC~4643, there seems
to be a point where the slope changes again corresponding to
the radius around the point of maximum ellipticity, which is not
considered as the end of the bar in the ellipticity-PA analysis. NGC~1433
shows an Fe3n distribution also compatible with no radial
change in the Fe3n, but there is a clear difference between the inner part and
the bar region. The Balmer indices show a positive gradient and a
change compared to the inner region. For two cases (NGC~2665 and NCG~3081)
there seems to be a negative trend of the metal indices and a slightly
negative trend in the Balmer indices; they do not show a change in
the slope in the bar region; interestingly enough these are the only two
galaxies classified as SAB in the sample, and they present an
exponential profile in the bar, which might indicate a different
formation process and history compared to those with a flat
profile~\citep{combes,noguchi}. This point deserves further
investigation. For all galaxies, there is also good agreement between
both sides of the bar.

There is an obvious influence of the different components to the shape
of the radial distribution. We have made tests subtracting a bulge
component (from the decomposition of the light profile), this
only helps to obtain an even clearer result for the bar and inner structure,
without changing the basic trends. We prefer to show the complete
radial profile in order to avoid affecting the the observed profiles with 
any model-dependent analysis.

Independently of the selected models (see Section~\ref{indices}), we
derive positive age and metallicity gradients in the bars (outer parts
of the  bars being younger and more metal rich than the internal
parts), except for the two galaxies with exponential bars (NGC~2665
and NGC~3081). In a future paper we plan to do a more detailed
analysis of the star formation history and chemical evolution of
all the objects. There, we will present the comparison of these results
with dynamical and stellar population models using the entire
spectra to try to interpret the results in the context of bar
formation and evolution.

Summarising, we have obtained high S/N long-slit spectra along the
bars of six early-type barred galaxies and we have derived radial
profiles of the line-strength indices for the sample galaxies. There
is a clear difference between the indices in the bar region (in
Fig.~\ref{fig:indices}, the region between the two vertical lines) and
the central component. Within the bar region there is a clear gradient
in both metal and Balmer lines that when combined with stellar
population models for all galaxies gives a gradient in both,
metallicity and age, indicating that outer parts of the bar are
younger and more metal rich. These results might indicate a secular
inside-out growth for the bar.

\begin{acknowledgements}

We are greatful to Reynier Peletier for allowing us to compare with
his code the emission correction. We are really greatful to B. Gibson
and K. Ganda for the careful reading of the manuscript. We thank the referee,
A. Vazdekis, for the very useful discussions and comments on the
manuscript. This publication makes use of data products from the Two
Micron All Sky Survey, which is a joint project of the University of
Massachusetts and the Infrared Processing and Analysis
Center/California Institute of Technology, funded by the National
Aeronautics and Space Administration and the National
Science. I.P. acknowledges financial support from the Netherlands
Organisation for Scientific Research (NWO) Foundation and the Leids
Kerkhoven-Bosscha Fonds. A.~Z. acknowledges support from  the
Consejer\'\i a de Educaci\'on y Ciencia de la Junta de Andaluc\'\i a,
Spain.

\end{acknowledgements}

\bibliography{ref}
\bibliographystyle{natbib}

\end{document}